# WHO DECIDES WHAT IS HARMFUL? CONTENT MODERATION POLICY THROUGH A MULTI-AGENT PERSONALISED INFERENCE FRAMEWORK

*Completed Research Paper*

Ewelina Gajewska, Warsaw University of Technology, Warsaw, Poland, ewelina.gajewska.dokt@pw.edu.pl

Michał Wawer, Warsaw University of Technology, Warsaw, Poland michal.wawer.stud@pw.edu.pl

Katarzyna Budzynska, Warsaw University of Technology, Warsaw, Poland, katarzyna.budzynska@pw.edu.pl

Jarosław A. Chudziak, Warsaw University of Technology, Warsaw, Poland, jaroslaw.chudziak@pw.edu.pl

## Abstract

*The increasing scale and complexity of online platforms raises critical policy questions around harmful content, digital well-being, and user autonomy. Traditional content moderation systems rely on centralised, top-down rules, often failing to accommodate the subjective nature of harm perception. This paper proposes an LLM-based multi-agent personalised inference framework that filters content based on unique sensitivity profiles of individual users. Our architecture combines domain-specific Expert Agents, a Manager Agent for orchestrating content analysis and agent selection, and a Ghost Profile Agent for simulating user perspectives, to inform moderation decisions. Evaluated against a range of non-personalised baselines, the system demonstrates up to a 32% improvement in accuracy, showing increased alignment with individual user sensitivities. Beyond technical performance, our framework provides policy-relevant insights for platform governance, providing a scalable way to reconcile moderation policies with societal and individual digital rights.*

*Keywords: Moderation Policy, Platform Governance, Harmful Speech, Large Language Models, Multi-Agent Systems*

## 1 Introduction

The proliferation of online platforms has led to an unprecedented scale of information exchange, where individuals continuously navigate a complex and often volatile digital public sphere. Automated content moderation has emerged as a dominant mechanism employed by platforms to manage harmful content such as hate speech, misinformation, and abusive language, aiming to protect users and uphold community standards (Gillespie, 2020). These platform-driven moderation systems typically operate with centralised, top-down rules that prioritise scalable enforcement and policy compliance over individual user needs and preferences (Gorwa et al., 2020). As a result, this approach often falls short in a politically charged environment where content can be highly contextual, and where users' diverse cultural, social, and ideological backgrounds challenge rigid moderation criteria (Gajewska et al., 2026; Kocoń et al., 2021; Sap et al., 2022). While some content may be broadly recognised as harmful, other material might be personally overwhelming, emotionally draining, or simply irrelevant, yet still impactful on mental well-being. To this end, users increasingly express a desire to control their content exposure beyond coarse categorisations of harm, reflecting a growing demand for agency in managing their online experience (Jin et al., 2017; Heung et al., 2025).

One of the central challenges in automated content moderation, mirroring similar debates in recommender systems and affective computing, is therefore the question of *what constitutes ground truth*. Unlike objective classification tasks, such as part-of-speech tagging, labelling harmful content is inherently subjective, shaped by individual moral frameworks, cultural norms, and social context (Cabitza et al., 2023; Fleisig et al., 2024; Sap et al., 2022). The same utterance may be perceived as hateful, humorous or acceptable depending on who interprets it. By defining ground truth in this perspectivist manner, we explicitly recognise that moderation systems must mediate between multiple, potentially conflicting interpretations of harmful content. Despite advances in recommender systems designed to personalise content delivery (Ekstrand et al., 2018), these systems principally optimise for user engagement metrics, sometimes at odds with users' long-term well-being or information diversity (Lasser & Poechhacker, 2025). Research has begun to reveal the ethical conflicts in maximising platform engagement that may reinforce echo chambers or exacerbate emotional distress. Yet, the dimension of user-driven filtering, where individuals exert granular control over the content they see, grounded in their subjective sensitivities rather than platform-defined harm, remains underexplored.

This paper addresses this critical gap by proposing a novel user-in-the-loop framework that empowers individuals to define and evolve personalised filters based on explicit feedback about the content they find overwhelming or harmful to their well-being. By capturing and inferring individual "sensitivity profiles", our system provides transparent and adaptive filtering that places the locus of control on the user rather than the platform. This approach responds to key open questions in the field, such as: **Q1.** How can personal preferences and sensitivities be modelled algorithmically to support dynamic filtering? **Q2.** How does shifting the locus of control from the platform to the user redefine the governance mechanisms of digital content moderation? And **Q3.** What system architectures best operationalise this user-centred control while maintaining baseline platform safety? By centring the user's agency and providing transparent mechanisms for content control, our work advances the understanding of hybrid, human-centric content moderation solutions that bridge the gap between platform governance and individual empowerment.

## 2 Related Work

Research on automated content moderation has long recognised the difficulty of defining and detecting harmful content in a way that is both user-dependent and socially inclusive. Early approaches relied heavily on lexical toxicity cues, later evolving toward contextual language models such as BERT that improved robustness but still struggled with ambiguity, cultural variability, and identity-related harms (Zampieri et al., 2020; Sap et al., 2022). Scholars have argued that such monolithic classifiers reflect and reinforce dominant social norms, often failing to capture the nuances of marginalised perspectives (Gillespie, 2020; Gorwa et al., 2020). Recent work highlights the subjectivity inherent in harmfulness judgments and the ethical limitations of one-size-fits-all moderation pipelines (Cabitza et al., 2023; Fleisig et al., 2024). To address the complex challenge of personalised hate speech moderation, our work draws upon two distinct but complementary streams of literature: personalised recommender systems and the perspectivist turn in computational linguistics.

While the proposed system utilises user-profiling techniques common in personalised recommender systems, the moderation of hate speech introduces a dual-constraint requirement: the system must simultaneously uphold platform-wide safety standards while respecting distinct user thresholds for offence. Traditional recommender systems architectures, often optimised for engagement via collaborative filtering (Bozdag, 2013), lack governance affordances to separate these distinct normative layers (Chen & Huang, 2024). Moreover, this optimisation can contribute to filter bubbles and echo chambers as algorithms repeatedly expose users to content that aligns with prior engagement patterns (Pariser, 2011). Recent studies highlight the tension between maximising engagement and supporting a healthy information ecosystem, underlining the ethical challenges of recommender system design (Helberger et al., 2021) and advocating for designs promoting agency over information consumption (Inkpen et al., 2019; Jhaver et al., 2023). These perspectives emphasise the importance of transparency, control, and respect for diverse values in algorithmic systems (Eslami et al., 2015). Here, interactive filtering interfaces that involve users in the decision process can increase trust and perceived fairness (Tsai & Brusilovsky, 2021).

However, operationalising these principles into scalable, real-time filtering designs that adapt to individual preferences remains challenging. User-in-the-loop frameworks for content moderation have emerged as promising approaches to balance automation with human judgment (Jhaver et al., 2022; Unsvåg & Gambäck, 2018). Yet, most such systems rely on predefined categories of harm or moderation policies rather than user-specific sensitivities that evolve over time (Malvicini et al., 2026). Moreover, current approaches often treat user feedback as a corrective mechanism rather than as a primary source of ground truth, thereby underutilising the diversity of user perspectives (Kuo et al., 2023). Consequently, there remains a need for user-adaptive models that can dynamically capture and learn from heterogeneity in users' perceptions of hate speech and online harm.

Recent advances in LLM architectures have demonstrated that multi-agent systems can enhance reasoning capabilities through collaborative inference (Uberna et al., 2026). The Mixture-of-Agents (MoA) approach (Wang et al., 2024) shows that aggregating responses from multiple LLMs through iterative refinement significantly improves performance compared to single-model approaches. Their layered architecture, employing proposer and aggregator agents, parallels our design where specialised expert agents provide domain-specific analyses aggregated by a Synthesis Agent. Since the effectiveness of such systems depends critically on context engineering (Gajewska et al., 2025; Zhang et al., 2025), our system in addition incorporates user-specific context to guide agent deliberation. Understanding user perspectives, however, requires inferring mental states and intentions—capabilities that remain challenging for LLMs. Research on Theory of Mind in LLMs finds that while models demonstrate surface-level perspective-taking, they struggle with deeper inferences about beliefs and emotional states (Chen et al., 2025; Wu et al., 2025). These limitations motivate our Ghost Profile Agent, which simulates user perspectives by explicitly encoding learned sensitivity thresholds rather than relying on implicit Theory of Mind capabilities. Furthermore, since structured prompting significantly improves LLM controllability in dialogue systems (Wagner & Ultes, 2024), we implement it in our approach, making the Manager Agent employ structured templates and combine base personas with dynamic, user-specific constraints.

Building on these architectural foundations, recent research has incorporated multi-agent architectures into hate speech detection to capture the contextual and dialogical nature of online harm. For instance, Park et al. (2024) propose PREDICT, a debate-style framework where agents adopt contrasting viewpoints, though their agents represent argumentative stances rather than user-centred sensitivities. Then, Masud et al. (2024) emphasise geographical and cultural context, showing that location-specific cues can meaningfully shift harm assessments. While these approaches advance socially informed moderation through multi-agent collaboration, they model context at a macro level rather than capturing individual user variation, falling short of genuinely adaptive filtering. Our work synthesises these strands (multi-agent collaboration, dynamic context engineering, and user perspective simulation), proposing an integrated system architecture for user-driven content filtering that leverages explicit user choices to infer dynamic sensitivity profiles, foregrounding user agency while maintaining scalable automation.

# 3 Approach

We propose a novel user-in-the-loop methodology to build a personalised content filtering system. The core of PRISM (**P**ersonalised **R**easoning and **I**nference **S**ystem for **M**oderation) approach is to infer a user's individual "sensitivity profile" based on their explicit feedback on hateful sentences. This profile is then used to filter future content, aiming to improve the user's digital well-being without relying on a pre-defined, one-size-fits-all model of what constitutes hateful content. Instead, the PRISM system dynamically adapts to each user's evolving perceptions and tolerance thresholds, allowing for a more user-dependent moderation process. By integrating user feedback loops, the model continuously refines its understanding of what the individual deems hateful or acceptable, thus reducing both over-censorship and under-moderation. This approach emphasises user agency and transparency, aligning content moderation practices with principles of participatory design (Vaccaro et al., 2021). Ultimately, the proposed framework seeks to balance freedom of expression with psychological comfort, fostering a healthier and more inclusive digital environment.

The system architecture consists of several coordinated modules structured for user-centred content moderation (see Figure 1). Specifically, it comprises three main stages: data preparation, a multi-agent inference engine, and a feedback module. The system architecture and evaluation framework are described in detail below.

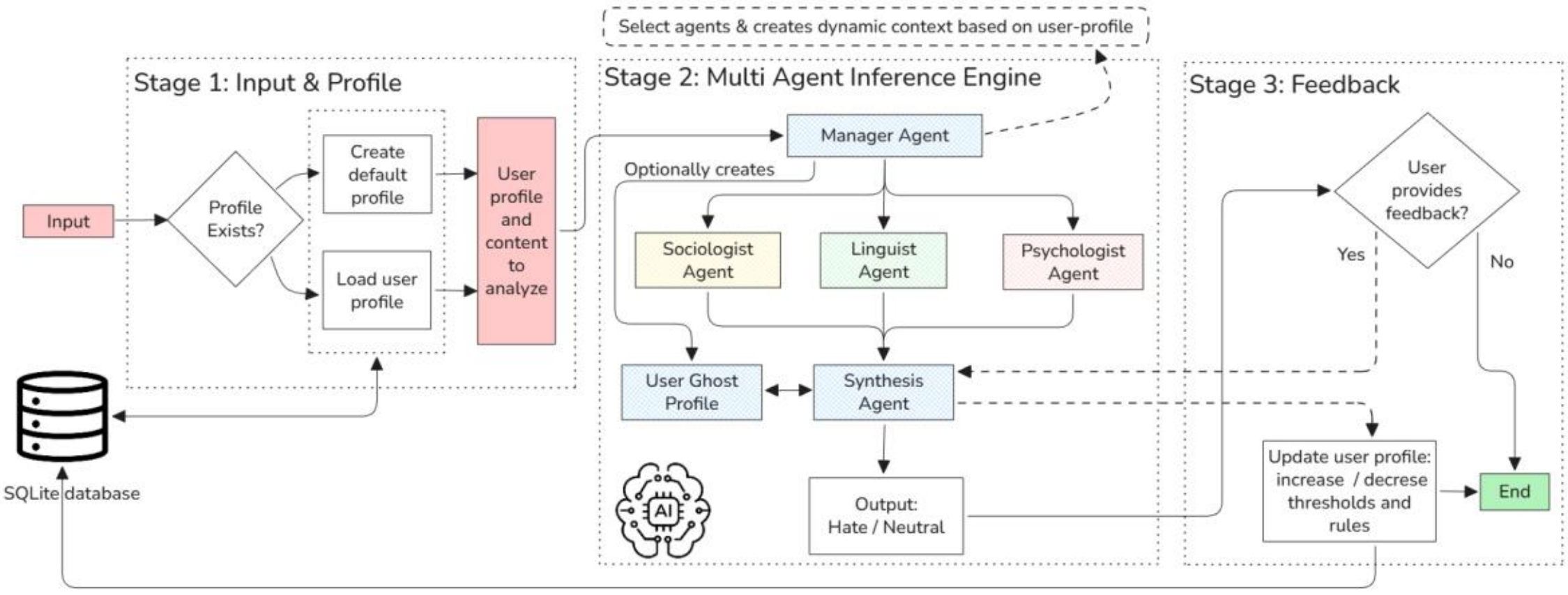

*Figure 1. PRISM architecture for personalised hate speech moderation. The system loads user-specific sensitivity profiles, employs specialised agents for multi-dimensional analysis, and synthesises personalised filtering decisions. User feedback enables continuous profile adaptation.*

## 3.1 User Profile Construction and Management

Each user profile $P_u$ comprises three core components that capture the user's personalised sensitivity across multiple dimensions of harmful content:

$$P_u = \begin{cases} user\ ID: u, \\ thresholds: \{d: t_u^{(d)} \mid d \in D\}, \\ weights: \{d: w_u^{(d)} \mid d \in D\}, \\ confidence: \{d: k_u^{(d)} \mid d \in D\}, \\ samples: n_u \end{cases}$$

Where $D$ = {sentiment, respect, insult, humiliate, status, dehumanise, violence, genocide, attack-defend, toxicity} represents dimensions of potentially harmful content derived from established hate speech taxonomies (Zampieri et al., 2020; Sachdeva et al., 2022). Unlike traditional binary classification systems that apply universal thresholds, our system maintains dimension-specific thresholds that reflect each user's individual tolerance levels. A user highly sensitive to violent content but tolerant of strong political rhetoric will have $t_u^{(violence)} < t_u^{(respect)}$ enabling personalised filtering aligned with their specific needs.

New users begin with no behavioural history, in the absence of any user information, the system initialises with population-level median. As users interact with the system, they provide explicit feedback by flagging content they find offensive or overwhelming. Each flagging action generates a training signal:

$$feedback_t = \{c_t, y_t, \{s_{c_t, d} \mid d \in D|\}\}$$

Where $c_t$ is the content at time $t$, $y_t \in \{hate, neutral\}$ is the user binary decision, and $s_{c_t, d}$ are the expert-generated severity scores for each dimension.

Dimension weights $w_u^{(d)}$ reflect the importance of each dimension in the user's filtering decisions. High weights indicate dimensions where the user demonstrates strong, consistent preferences; low weights suggest dimensions of lesser concern. Weights are computed from the variance of severity scores across all content the user has seen:

$$w_u^{(d)} = \sigma\left(\{s_{c, d} \mid c \in C_u\}\right)$$

Where $c_u$ is the complete set of content shown to user $u$ and $\sigma$ denotes standard deviation. Intuitively, dimensions where content severity varies widely (high $\sigma$ ) are more discriminative for filtering decisions and receive higher weights.

Profile confidence quantifies the reliability of learned thresholds and weights based on the volume of user feedback:

$$\kappa_u^{(d)} = \min\left(\frac{n_u}{100}, 1\right)$$

Where $n_u = |c_u|$ is the total number of content items the user has rated (flagged or shown). Confidence asymptotically approaches unity as the user provides more feedback, achieving full confidence (k = 1) at 100 interactions. After internal testing, we set this number to 100 because it allows all dimensions to stabilise, and at around this number, the system achieves stable results. Of course, this number may vary for different users. Confidence governs the system's reliance on personalised versus population-level information. New users ($\kappa_u \rightarrow 0$) default to population priors while experienced users ($\kappa_u \rightarrow 1$) rely on learned preferences. Confidence is also communicated to expert agents via the Manager's dynamic context (e.g., "Profile confidence: 0.35 — weight population norms in ambiguous cases"), allowing agents to adjust reasoning accordingly.

## 3.2 Multi-Agent Inference Engine

Multi-agent architecture mimics human deliberative judgment by decomposing content analysis into specialised expert perspectives. This design addresses a fundamental limitation of monolithic classifiers: the inability to capture the multidimensional, user-dependent nature of hate speech perception. Our MAS consist of following agents:

**Manager Agent** serves as the orchestrator of the analysis pipeline. Upon receiving content and a user profile, it performs three functions:

1. Content Analysis: Examines the text to identify salient features such as identity-targeting language, aggressive rhetoric, or threatening content.
2. Expert Selection: Dynamically determines which expert agents are relevant for the given content. Rather than invoking all experts uniformly, the manager selectively activates 1-3 specialists based on content characteristics.
3. Context Engineering: Generates dynamic, personalised prompt fragments for each selected expert. These user-dependent context fragments guide experts to focus on aspects most relevant to both the content and the user's sensitivity profile.

For instance, the Manager might generate the following context for the Sociologist Agent:

*This user is highly sensitive to identity attacks (threshold: 0.2). Focus on implicit othering and dehumanising comparisons. Note: user threshold for dehumanisation is 0.15-extremely low tolerance.*

To ensure consistent LLM interpretation, the Manager Agent maps raw parameters to calibrated natural language descriptors before constructing expert prompts. Thresholds are translated to sensitivity levels (e.g., below 0.15: "extremely sensitive"; 0.55–0.75: "tolerant") and weights to importance descriptors (e.g., above 0.8: "primary concern"; below 0.2: "negligible"). Each **expert** prompt therefore pairs the numerical value with its semantic anchor and provide domain-specific analysis:

1. Sociologist Agent:
   - Expertise: Identity-based discrimination, intersectionality, systemic oppression, dehumanisation patterns.
   - Analysis Lens: Examines content through sociological frameworks to identify targeting of protected groups, stereotypes, and marginalisation tactics.
2. Linguist Agent:
   - Expertise: Rhetorical patterns in harmful speech, toxic language markers, linguistic aggression.
   - Analysis Lens: Analyses linguistic structures, aggressive rhetoric, insults, and disrespectful language patterns.
3. Psychologist Agent:
   - Expertise: Psychological impact of online harm, emotional violence, threat perception.
   - Analysis Lens: Evaluates threats, fear-inducing language, and potential for psychological distress.

Each expert receives a composite prompt combining three components: *BASE PROMPT + DYNAMIC CONTEXT + TASK*. The base prompt defines the expert's static persona, the dynamic context provides personalised guidance from the Manager, and the task template specifies the standard analysis task. Complete prompt templates for all agents, including base personas, dynamic context generation logic, and task specifications, are available in the supplementary repository[1].

**Ghost Agent** is dynamically created when the Manager determines that personalised perspective simulation would enhance accuracy, it invokes the Ghost Profile Agent. This agent embodies the specific user's sensitivity profile, making judgments as the user themselves might. Unlike the expert agents that provide domain-specific analysis, the Ghost Profile directly simulates user perception based on learned thresholds and sensitivities. The Manager invokes the Ghost Profile when multiple experts disagree, requiring a tie-breaker aligned with user preferences.

**Synthesis Agent** aggregates all expert analyses into a final decision. It (1) Gathers decisions, reasoning, and confidence scores from all invoked agents, (2) Considers areas of agreement/disagreement and confidence levels and (3) Produces final binary classification (hate/neutral).

## 3.3 Learning and Profile Updates

User preferences evolve over time due to changing contexts, emotional states, or shifting concerns. To maintain alignment, the system continuously updates profiles through online learning using exponential moving average (EMA) - as presented in the algorithm below. The Synthesis Agent receives feedback signal and invokes tools to update user profile. The system employs an adaptive learning rate (α) that dynamically adjusts based on the accumulated sample count and user confidence, calculated as 0.1 + 0.2 (1-$\kappa_u$), where $\kappa_u$ represents the user's confidence score capped at 1.0. For content flagged as offensive, the algorithm iterates through each dimension and lowers the corresponding threshold by computing the weighted average between the current threshold and the observed severity score, effectively increasing the user's sensitivity to similar content. Conversely, when users choose not to flag content, thresholds are raised if the observed severity exceeds current expectations by a tolerance factor $\delta = 0.1$, making the filter more permissive. The Synthesis Agent then utilises its database communication tools to persist the updated profile to the SQLite database.

---

[1] https://github.com/michal-wawer/PRISM

**Require:** Profile $\mathcal{P}_u$, feedback $(c_t, y_t, \{s_{c_t,d}\})$
**Ensure:** Updated profile $\mathcal{P}'_u$
1: $n_u \leftarrow n_u + 1$ ▷ Increment sample count
2: $\kappa_u \leftarrow \min\left(\frac{n_u}{100}, 1\right)$ ▷ Update confidence
3: $\alpha \leftarrow 0.1 + 0.2 \cdot (1 - \kappa_u)$ ▷ Adaptive learning rate
4: **if** $y_t =$ hide **then** ▷ User flagged content
5: **for** each dimension $d \in \mathcal{D}$ **do**
6: $s_{\text{obs}} \leftarrow s_{c_t,d}$ ▷ Observed severity
7: $t_{\text{curr}} \leftarrow t_u^{(d)}$ ▷ Current threshold
8: **if** $s_{\text{obs}} < t_{\text{curr}}$ **then** ▷ User more sensitive than expected
9: $t_u^{(d)} \leftarrow (1 - \alpha) \cdot t_{\text{curr}} + \alpha \cdot s_{\text{obs}}$ ▷ Lower threshold
10: **end if**
11: **end for**
12: **else** ▷ User did not flag content
13: **for** each dimension $d \in \mathcal{D}$ **do**
14: $s_{\text{obs}} \leftarrow s_{c_t,d}$
15: $t_{\text{curr}} \leftarrow t_u^{(d)}$
16: **if** $s_{\text{obs}} > t_{\text{curr}} + \delta$ **then** ▷ User more tolerant than expected, $\delta = 0.1$
17: $t_u^{(d)} \leftarrow (1 - \alpha) \cdot t_{\text{curr}} + \alpha \cdot s_{\text{obs}}$ ▷ Raise threshold
18: **end if**
19: **end for**
20: **end if**
21: Recompute weights $\{w_u^{(d)}\}$ using Equations 7–14 **return** $\mathcal{P}_u$

Figure 2. Algorithmic representation of a adaptive learning.

## 3.4 Content Filtering Workflow

When new content $c$ arrives for user $j$ :

1. Load user's current sensitivity profile $P_j$ from SQLite database
2. Manager agent orchestrates expert evaluation, producing scores $\left\{s_i^{(d)},\ i \in agents\ ,\ d \in D\right\}$
3. Synthesis agent generates weighted decision score
4. Compare score against user's personalised threshold
5. Hide content if score exceeds threshold; otherwise display

# 4 Evaluation Framework

Our experimental design investigates whether personalised content filtering can demonstrably outperform universal moderation approaches (RQ1), and whether the proposed multi-agent architecture offers advantages over alternative personalisation strategies (RQ2 and RQ3). In our experiments we used GPT-4.1-mini model to balance performance and cost of our experiments.

## 4.1 Dataset

To evaluate the proposed system, we employ the Measuring Hate Speech dataset (Sachdeva et al., 2022), a large-scale corpus of social media comments. It was selected due to its comprehensive annotation scheme, which captures the multidimensional nature of offensive content through both fine-grained constituent labels and a continuous hate speech score derived via Item Response Theory (IRT) (Eckes, 2023). Importantly, the dataset also includes disaggregated annotation data, meaning that individual annotator judgments are preserved rather than aggregated into a single consensus label. This allows us to model subjective variation in the perception of hate speech, reflecting how different users may disagree on what constitutes offensive content. Leveraging these individual-level annotations, we construct separate "ground truth" for distinct users, each representing a unique tolerance threshold. This structure enables the development and evaluation of a user-centred frameworks under conditions that more closely mirror real-world diversity in user feedback. Consequently, the Measuring Hate Speech dataset provides not only rich linguistic and topical information but also the necessary granularity to support the personalised and dynamic aspects of our approach.

The dataset comprises 39,565 unique comments collected from social media platforms, annotated by 7,912 distinct users, yielding a total of 135,556 annotation instances. This multi-user design enables the dataset to capture the inherent subjectivity and perspectivism in hate speech perception. Each comment in the dataset is accompanied by annotations across multiple dimensions. The primary outcome variable is a continuous hate speech score, where higher values indicate greater perceived hatefulness. The scoring convention defines values exceeding 0.5 as indicative of hate speech, values below -1.0 as counter-speech or supportive content, and intermediate values (-1.0 to 0.5) as neutral or ambiguous. This continuous measure is derived from ten constituent ordinal labels: sentiment, respect, insult, humiliation, status, dehumanisation, violence, genocide, attack/defence orientation, and a binary hate speech indicator. The aggregation of these dimensions through IRT modelling accounts for variation in annotator interpretation and provides a more nuanced representation of content offensiveness than traditional binary classification schemes.

## 4.2 Experiment 1: Universal vs. Personalised Filtering

To investigate whether personalised content filtering outperforms universal, one-size-fits-all moderation across different sensitivity dimensions, we hypothesise that personalised models will achieve better alignment with individual annotator preferences than universal models, particularly for users with extreme sensitivity profiles (very strict or very lenient). We test this hypothesis by comparing two filtering approaches: (1) *a Universal Filter* serving as the baseline condition (calculated as mean of all 100 annotator profiles), which applies a single, global decision rule across all users, specifically, flagging comments with a hate speech score exceeding 0.5 while retaining those below this threshold. This configuration emulates conventional platform moderation strategies employed by large social media services (e.g., Facebook), where moderation policies are derived from aggregated judgments and applied uniformly regardless of individual tolerance levels. In contrast, (2) A Single-Agent Personalised filter, which receives the same user-specific sensitivity profile as the full system but consolidates all reasoning into a single GPT-4.1-mini call without expert decomposition, dynamic context engineering, or synthesis-based aggregation. This condition isolates the contribution of personalisation from the contribution of the multi-agent architecture (3) *the Personalised Filter* assigns each synthetic user a distinct sensitivity profile, defined by user-specific thresholds on primary content dimensions. Each profile captures a particular evaluative focus (e.g., emotional tone or respectfulness) and reflects the variability observed among individual users in the Measuring Hate Speech dataset. Thresholds for each dimension are learned empirically from individual-level user annotation data, ensuring that the personalised filters approximate the real decision boundaries observed in human judgments. Through this setup, Experiment 1 contrasts the effectiveness of a static, universal moderation policy against dynamically tuned, user-specific models, allowing us to assess whether the proposed user profiling module yields measurable improvements in alignment with user-level preferences and sensitivities.

### 4.2.1 Profile Selection

Each user in the dataset provides a series of annotation decisions indicating their assessment of potentially hateful material. However, as illustrated in Figure 3, the number of annotation decisions contributed by individual users is relatively low. This unbalanced distribution poses challenges for conducting analyses at the individual level, as the limited number of decisions per user reduces statistical reliability and increases the potential influence of noise. To address this issue, we aggregate data into 100 user profiles based on three primary criteria designed to ensure both statistical reliability and representativeness across the population of annotators. (1) *Annotation volume*: annotators with $n_i \in [10, 25]$, where $n_i$ denotes the annotation count for annotator *i.* This range captures the modal region of the distribution while ensuring sufficient data for profile estimation. (2) *Severity diversity*: Stratified sampling across five severity categories defined by the annotator severity parameter $\alpha_j$ from the faceted partial credit IRT model (Eckes, 2023; Kennedy et al., 2020):

$$S_j = \begin{cases} very\ strict\ if\ \alpha_j < -1.0 \\ strict\ if -1.0 \leq \alpha_j < -0.3 \\ moderate\ if -0.3 \leq \alpha_j < 0.3 \\ lenient\ if\ 0.3 \leq \alpha_j < 1.0 \\ very\ lenient\ if\ \alpha_j \geq 1.0 \end{cases}$$

(3) *Demographic diversity*: Within each severity stratum, we sample to maximise heterogeneity across age, race, gender, and political ideology, though this was treated as a secondary criterion given the primary importance of severity diversity for evaluating personalisation effectiveness.

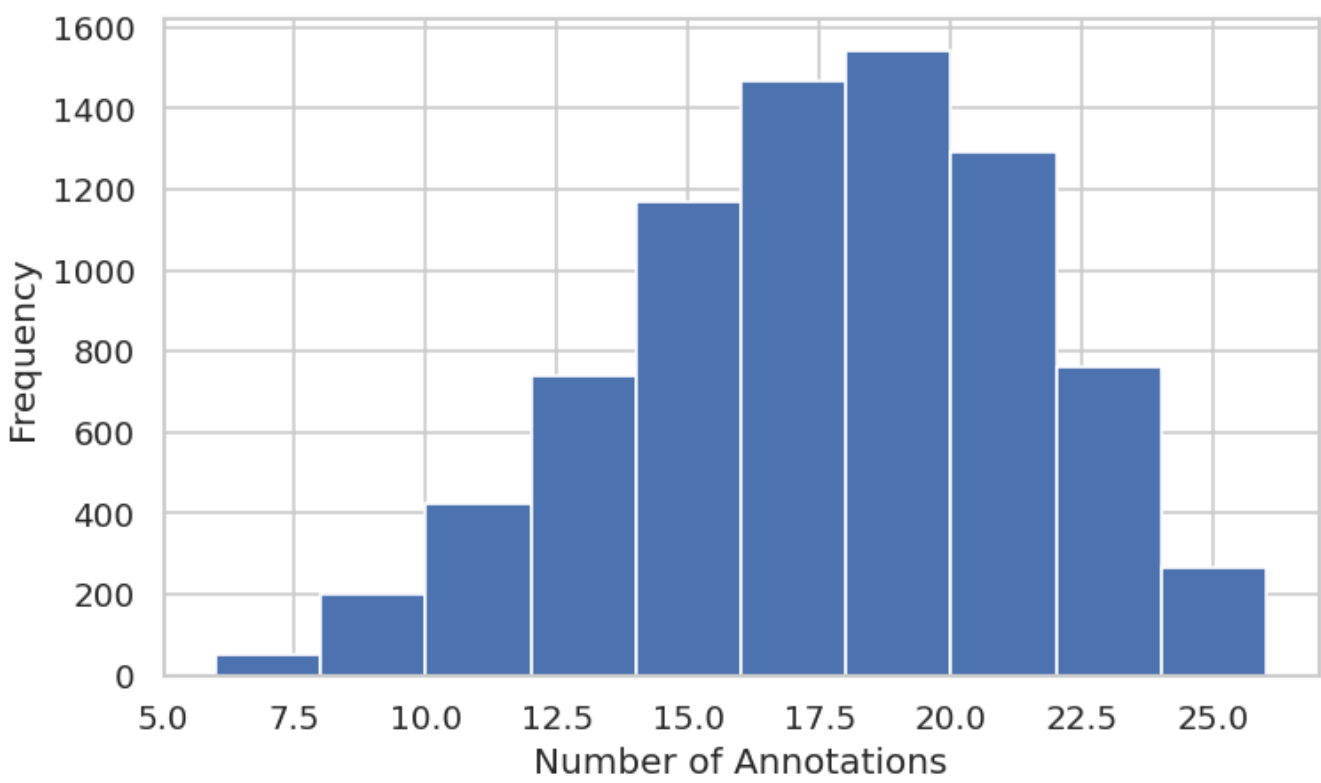

*Figure 3. Number of users and the number of their annotation decisions in the dataset.*

This aggregation enables us to capture broader patterns in annotation behaviour while mitigating sparsity issues caused by limited contributions from single users. In this way, we preserve variation between groups while ensuring that each profile contains a sufficient number of annotation decisions for rigorous quantitative analysis. The annotator severity parameter $\alpha_j$ is estimated through the faceted partial credit model:

$$\log\left[\frac{p_{nijk}}{p_{nijk-1}}\right] = \theta_n - \delta_i - \alpha_j - \tau_k$$

where $\theta_n$ represents the latent severity of comment *n*, $\delta_i$ denotes the difficulty of rating scale item *i*, $\alpha_j$ captures the systematic bias (severity) of rater *j*, and $\tau_k$ represents the threshold between response categories *k* and *k-1*. Negative values of $\alpha_j$ indicate that annotator *j* systematically interprets content as more severe than the population mean (strict), while positive values indicate more lenient interpretations.

### 4.2.2 Profile Construction Methodology

For each selected annotator *j*, we constructed a profile $P_j$ comprising dimension-specific thresholds, importance weights, and confidence scores. The construction process operates on the ordinal ratings provided by annotators rather than the continuous IRT-derived scores, preserving the direct relationship between annotator behaviour and profile parameters. Our sampling strategy aimed to select approximately 20 profiles from each severity category, ensuring adequate representation across the full spectrum of annotator perspectives. The procedure is presented in Figure 4. The sampling procedure successfully identified 100 annotator profiles meeting our criteria. The distribution of profiles across severity categories is presented in Table 2, showing a good balance across the spectrum.

| Severity Category | Count | Mean $\alpha_j$ |
|---|---|---|
| Very strict $(\alpha_j < -1)$ | 18 | -1.32 |
| Strict $(-1 \leq \alpha_j < -0.3)$ | 21 | -0.58 |
| Moderate $(-0.3 \leq \alpha_j < 0.3)$ | 24 | 0.02 |
| Lenient $(0.3 \leq \alpha_j < 1)$ | 20 | 0.61 |
| Very lenient $(\alpha_j \geq 1)$ | 17 | 1.28 |

*Table 2. Distribution of selected profiles across severity categories.*

1: **Input:** Dataset $\mathcal{D}$, target count $N = 100$, annotation range $[n_{\min}, n_{\max}] = [10, 25]$
2: **Output:** Set of selected annotator IDs $\mathcal{A}^*$
3: $\mathcal{E} \leftarrow \{j \mid n_j \in [n_{\min}, n_{\max}]\}$ ▷ Eligible annotators
4: **for** each severity category $s \in$ {very strict, strict, moderate, lenient, very lenient} **do**
5: $\quad \mathcal{E}_s \leftarrow \{j \in \mathcal{E} \mid S_j = s\}$ ▷ Filter by severity
6: $\quad k_s \leftarrow \min(N/5, |\mathcal{E}_s|)$ ▷ Target: 20 per category
7: $\quad \mathcal{A}_s^* \leftarrow \textsc{RandomSample}(\mathcal{E}_s, k_s)$ ▷ Stratified sample
8: $\quad \mathcal{A}^* \leftarrow \mathcal{A}^* \cup \mathcal{A}_s^*$
9: **end for**
10: **if** $|\mathcal{A}^*| < N$ **then**
11: $\quad \mathcal{E}' \leftarrow \mathcal{E} \setminus \mathcal{A}^*$ ▷ Remaining eligible
12: $\quad \mathcal{A}_{\text{add}} \leftarrow \textsc{RandomSample}(\mathcal{E}', N - |\mathcal{A}^*|)$
13: $\quad \mathcal{A}^* \leftarrow \mathcal{A}^* \cup \mathcal{A}_{\text{add}}$
14: **end if**
**return** $\mathcal{A}^*$

*Figure 4. Algorithmic representation of a stratified profile selection for personalised inference.*

Figure 5 presents a comprehensive visualisation of selected profiles across two dimensions: annotation count and annotator severity, with color encoding indicating severity category. This visualisation confirms that our selection achieved the desired diversity across both dimensions, with profiles distributed evenly across the severity spectrum while maintaining the target annotation range.

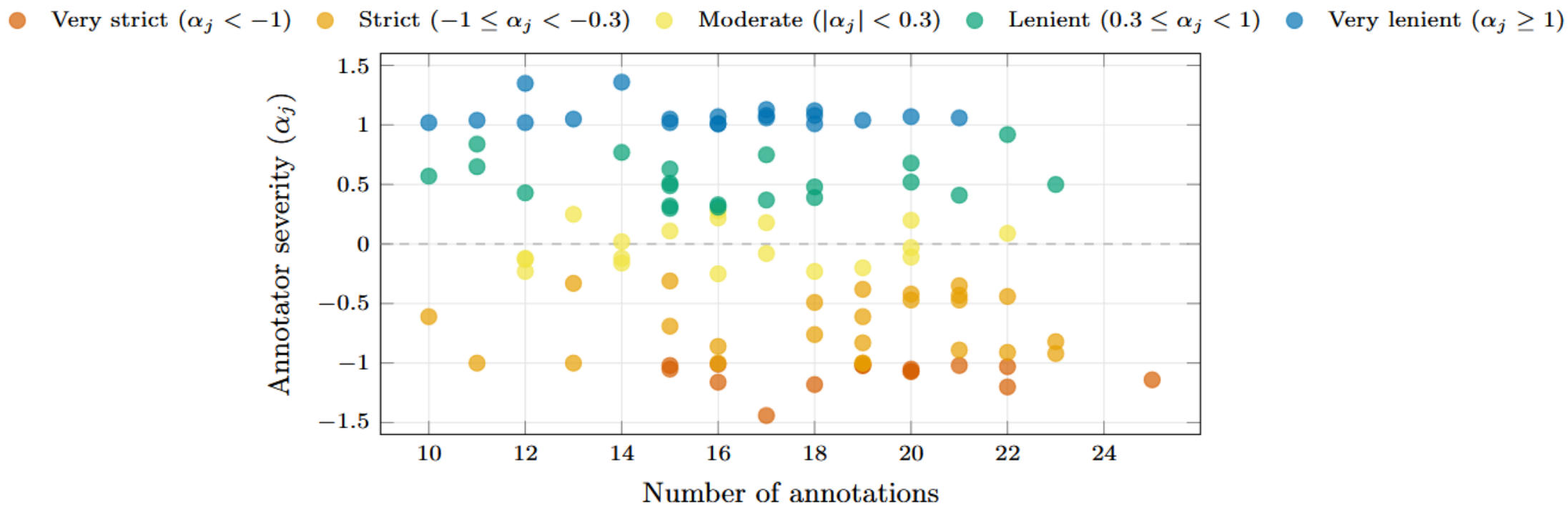

*Figure 5. Visualisation of the 100 extracted annotator profiles. Each point represents one annotator and color indicating severity category.*

### 4.2.3 Ground Truth Definition

In this study, we adopt a perspectivist approach to define what constitutes ground truth in the context of hate speech annotation. Traditional annotation paradigms typically assume the existence of a single, objective label that represents the true meaning or intent of a text. However, such an assumption is often problematic in subjective domains like hate speech detection (see Cabitza et al., 2023; Fleisig et al., 2024), where interpretation inherently depends on the annotator's personal, cultural, and social background (Gajewska et al., 2026; Sap et al., 2022). Following these recent perspectivist proposals, we treat each annotation as a valid expression of an individual's perspective rather than as noise or error relative to an assumed objective standard. Consequently, our operationalisation of ground truth reflects a plurality of perspectives rather than a singular, universal truth. This perspective-aware conception enables a richer understanding of hate speech as a socially constructed and contested phenomenon, aligning with ongoing shifts in computational social science toward more inclusive and reflexive annotation practices.

#### 4.2.4 Metrics

Model performance is evaluated using classification-based metrics to capture alignment between the system's predictions and individual user judgments. First, *F1* score measures the harmonic mean of precision and recall, providing a balanced indicator of the model's ability to correctly identify offensive content while minimising false positives and false negatives. To ensure robustness of observed differences between models, we conduct *paired t-tests* comparing performance results across experimental condition. Additionally, we compute Cohen's to estimate the effect size of pairwise model differences, providing an interpretable measure of the magnitude of improvement.

## 4.3 Experiment 2: Inference-based User Profiling vs. Alternative Architectures

To assess whether the proposed inference-based user profiling framework offers advantages over alternative architectures in terms of adaptability and alignment with user-specific preferences, we hypothesise that models incorporating individual annotator history will outperform universal, non-personalised baselines. Rather than reimplementing baseline systems, we benchmark our personalised multi-agent approach against state-of-the-art transformer-based classifiers tested on the same dataset, reported by Antypas and Camacho-Collados (2023), specifically leveraging their "All*" metric, which represents model performance when trained on a balanced sample across all datasets of the same size as the best single-dataset baseline. This cross-dataset training approach provides a generalisation-focused benchmark. We compare our system against four widely adopted models: BERTweet, TimeLMs, RoBERTa, and BERT-all. These baselines serve as static references for harmfulness detection, enabling direct comparison between standard non-personalised moderation and our user-adaptive multi-agent system. Performance is measured using the macro-averaged F1 score.

## 4.4 Experiment 3: Learning Dynamics

To investigate how classification performance evolves as a function of feedback volume, we conduct a learning curve analysis comparing the PRISM system against a single-agent and universal baseline. This experiment examines how quickly personalised profiles converge to stable performance. For each user profile $j$ with $n_j$ annotations, we construct a series of progressively enriched profiles by using only the first $k$ annotations for profile construction via EMA updates (Section 3.3), with the remaining $n_j - k$ annotations held out for evaluation. We vary $k \in [2, 22]$, excluding users with fewer than $k + 3$ annotations at each step to ensure proper evaluation sets. Since annotators in the Measuring Hate Speech dataset contributed between 10 and 25 annotations each (Section 4.1), this hold-out requirement constrains the maximum training budget to $k = 22$; the resulting learning curve therefore captures the early-to-mid adaptation regime rather than long-term convergence behaviour, which we identify as a priority for future evaluation.

# 5 Results

This section presents findings from our experiments, addressing whether personalised content filtering outperforms universal moderation and whether our multi-agent architecture offers advantages over existing state-of-the-art systems.

## 5.1 Experiment 1: Universal vs. Personalised Filtering

Results reported in Table 3 demonstrate that incorporating user-specific sensitivity profiles yields substantial improvements in classification performance across all evaluation metrics. The single-agent personalised system achieved a macro-averaged F1 of 74.5%, a relative improvement of 20.6% over the universal baseline (61.8%), confirming that user-specific sensitivity profiles alone substantially improve alignment with individual preferences. MAS-Personalised extends these gains further, achieving 81.5% macro-averaged F1, 7.0 pp improvement over the single-agent and a 31.9% relative improvement over the universal filter. Notably, the transition from single-agent to multi-agent yields disproportionate gains in precision (from 78.9% to 85.2%) compared to recall (from 79.4% to 83.5%), suggesting that expert decomposition primarily helps the system avoid false positives by enabling dimension-specific reasoning that a single model conflates. Cohen's Kappa improved from 0.58 to 0.64 (substantial agreement), indicating that multi-agent deliberation enhances not only aggregate accuracy but also consistency with individual user judgments.

Analysis of per-class performance reveals important characteristics of the personalised system's decision-making behaviour. For hate speech detection, the system achieved exceptionally high recall (99.73%), correctly identifying 750 of 752 instances, with only 2 false negatives. This demonstrates the system's ability to effectively protect users from harmful content aligned with their sensitivity profiles. Precision for hate speech was 70.69%, reflecting 311 false positives where neutral content was classified as offensive. From a user experience perspective, this performance profile suggests that users may occasionally encounter over-filtering of borderline content but will rarely be exposed to harmful material that exceeds their personalised sensitivity thresholds. This design philosophy reflects the priority of user protection and digital well-being over content accessibility, acknowledging that the psychological cost of exposure to unwanted harmful content typically exceeds the inconvenience of occasionally over-cautious filtering.

| **Approach** | **Macro F1 (%)** | **Precision (%)** | **Recall (%)** | **Cohen's Kappa (-1 to +1)** | **Support** |
|---|---|---|---|---|---|
| Universal (baseline) | 61.8 | 74.7 | 69.8 | 0.33 | 1702 |
| Single-Agent Personalised | 74.5 | 78.9 | 79.4 | 0.58 | 1702 |
| PRISM (MAS - Personalised) | 81.5* | 85.2* | 83.5* | 0.64* | 1702 |
| Relative improvement % | +31.9 | +14.0 | +19.6 | +49.5 | - |

*Table 3. Classification performance comparison between universal baseline and personalised filtering across 100 user profiles (n = 1,702 test instances). Both baseline and a personalised version used gpt-4.1-mini LLM. Asterisk indicates better performance.*

## 5.2 Experiment 2: Comparison with State-of-the-Art Baselines

Table 4 presents a comparative evaluation of our personalised multi-agent system against four widely adopted transformer-based architectures evaluated on the same dataset. The baseline performance values are drawn from Antypas and Camacho-Collados (2023). These baselines represent state-of-the-art non-personalised approaches to hate speech detection, providing a rigorous external benchmark for evaluating our system's effectiveness. Results demonstrate that our personalised multi-agent system (81.5% macro-averaged F1) outperforms all non-personalised transformer baselines. The closest competitor is BERTweet (80.5%), a Twitter-specific language model pre-trained on 850 million tweets, which our system exceeds by 1.0 percentage point (representing a 1.2% relative improvement). This advantage is particularly notable given that BERTweet benefits from domain-specific pre-training on social media data, suggesting that user-specific adaptation provides complementary value beyond architectural sophistication or domain specialisation. The performance gap widens substantially when compared to general-purpose transformers: our system outperforms BERT (75.0%) by 6.5 percentage points (+8.7% relative improvement), TimeLMs (74.9%) by 6.6 percentage points (+8.8%), and RoBERTa (73.7%) by 7.8 percentage points (+10.6%).

| **System** | **Macro-averaged F1** | **Personalised** |
|---|---|---|
| BERTweet | 80.5 | No |
| TimeLMs | 74.9 | No |
| RoBERTa | 73.7 | No |
| BERT | 75.0 | No |
| PRISM (Ours) | 81.5* | Yes* |

*Table 4.* *Macro-averaged F1 scores comparing our personalised multi-agent system against state-of-the-art transformer-based baselines. Asterisk indicates the best performance.*

## 5.3 Experiment 3: Learning dynamics

Figure 6 presents the learning curve for both the personalised multi-agent system and the single-agent personalised as a function of training interactions k. Both systems surpass the universal baseline (F1 = 0.618) from the earliest evaluation point (k = 2), indicating that profile build from average weights of the annotators is much worse than the default profile with little feedback. This result alleviates cold-start concerns: users benefit from personalised filtering immediately upon providing initial feedback, without an extended calibration period. Combined with the findings from Experiment 1, where multi-agent deliberation disproportionately improved precision (Section 5.1), this pattern indicates that expert decomposition becomes increasingly valuable as the profile provides more nuanced dimensional information for each agent to act upon.

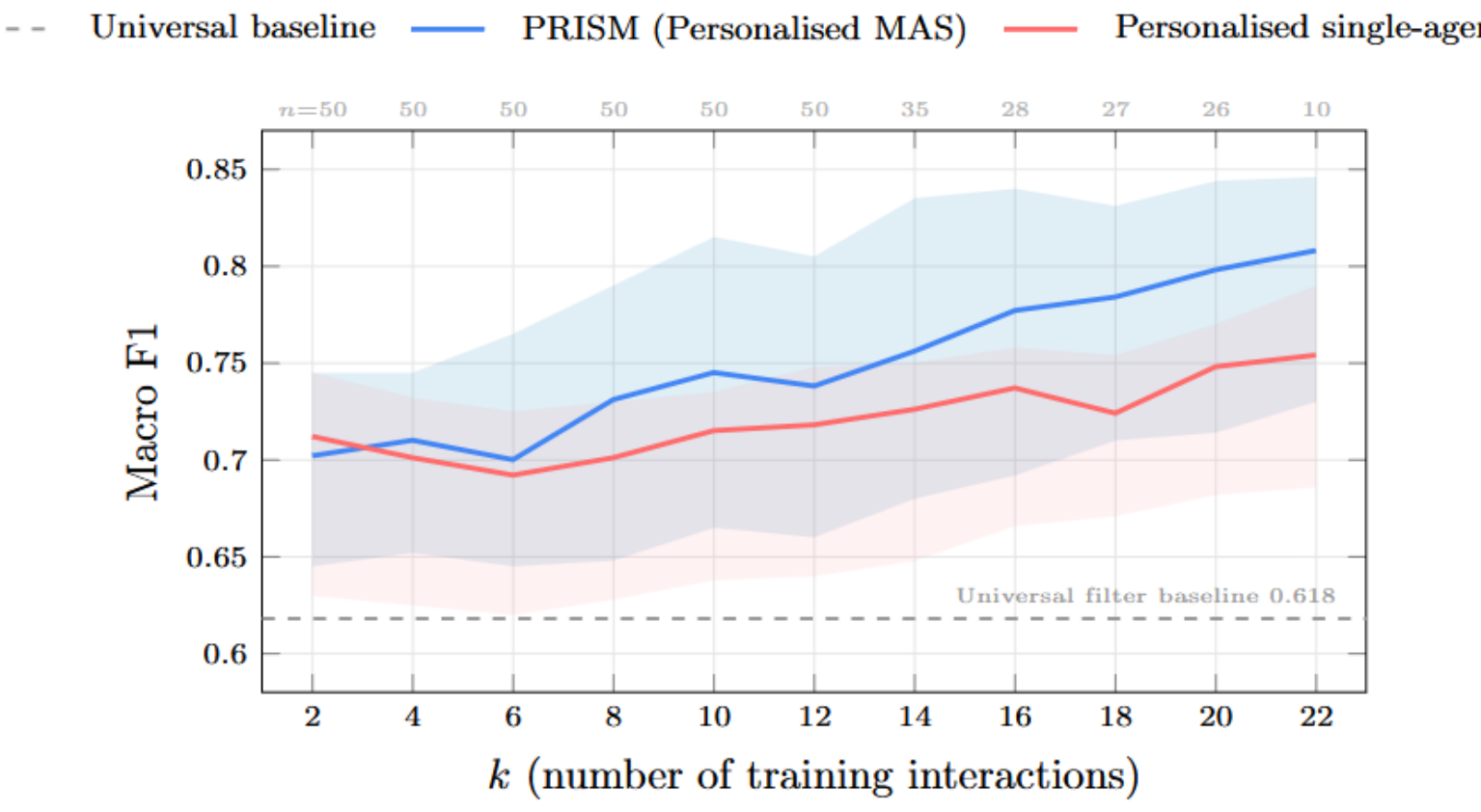

Figure 6. Learning curve comparing personalised MAS and baseline as a function of training interactions k (*n* along top indicates number of user profiles evaluated at each *k)*

# 6 Discussion

Traditional moderation systems rely on monolithic classifiers that attempt to learn a global decision boundary for harmful content. This approach has been repeatedly critiqued for flattening complex social and contextual signals, failing to capture marginalised perspectives, and reproducing biases embedded in training data (Gillespie, 2020; Hovy & Spruit, 2016; Sap et al., 2022). While personalisation has been widely explored in recommender systems (Ekstrand et al., 2018), its role in moderation has remained largely unexamined. Our design addresses these concerns by using personalisation to mediate disputes among agents and align decisions with explicit user sensitivity profiles. In doing so, we extend prior work on user-in-the-loop moderation tools (Jin et al., 2017) by embedding user agency directly into the inference pipeline. This architectural choice offers additional benefits beyond classification accuracy, including interpretability through agent-specific reasoning, transparency in decision-making, and the ability to directly incorporate user feedback for continuous adaptation. Consequently, the results support the feasibility of personalised, user-in-the-loop content moderation as both a technically effective and socially desirable approach to platform governance.

Implementing such a user-in-the-loop system, however, necessitates navigating complex policy trade-offs and value judgments. Foremost, it must balance autonomy vs. burden-shifting. While giving users control over their moderation filters enhances individual agency, it risks shifting the cognitive and emotional labour of safety from the platform onto the victim, potentially exacerbating the exhaustion already experienced by marginalised groups. To mitigate this tension, we propose two concrete design principles for platform governance that operationalise autonomy without overwhelming the user. First, the architecture must be designed so that the platform's overarching policy agent retains the primary burden of enforcing baseline safety standards. The system should operate on a principle of 'opt-out' for severe toxicity, ensuring that the user-in-the-loop mechanism is only activated for highly contextual edge cases rather than foundational safety. Second, platforms can facilitate collective agency by allowing users to subscribe to moderation profiles developed by trusted civil society organizations or community advocates. This delegates the heavy cognitive labour of defining and updating hate speech thresholds to specialised entities, preserving the individual user's autonomy to choose their moderation experience without forcing them to build it from scratch. It also prevents the traumatisation of harm associated with manual content review: it provides personalisation without forcing users to explicitly read and label every toxic comment to train their custom moderation filter.

# 7 Conclusions

This paper set out to address a foundational question in contemporary digital governance: *Who decides what is harmful?* By reframing harmfulness as a perspectivist, user-dependent construct rather than a universal property, we developed and empirically validated a multi-agent personalised inference framework for content moderation. Our results demonstrate that personalised moderation, grounded in user-specific sensitivities in addition to platform-wide rules, can be both technically feasible and socially impactful. Across two experiments, we showed that integrating user-specific information significantly enhances model performance relative to universal baselines. Experiment 1 revealed consistent and statistically significant improvements in F1 scores (+31.9%), with medium-to-large effect sizes and, critically, equitable gains across age, gender, and racial groups. Experiment 2 extended these findings by demonstrating that dynamic, continuously updated user profiles outperform static personalisation strategies, highlighting the importance of modelling evolving user behaviour in real-world deployments. Conceptually, we introduce a shift from platform-centric to user-centred moderation by anchoring decisions in subjective sensitivity profiles. This reconceptualisation challenges the long-standing assumption that harmfulness can and should be adjudicated through universal policies. Methodologically, we propose a multi-agent architecture that enables transparent, real-time adaptation to users' evolving preferences, bridging works on recommender systems, affective computing, and moderation policy in a novel way. Empirically, our experiments provide robust evidence that personalised inference offers stable performance improvements and fairness gains across demographic groups, addressing persistent concerns about representational biases in current moderation systems.

## 8 Acknowledgements

We would like to acknowledge that the work reported in this paper has been supported in part by the Polish National Science Centre, Poland (Chist-Era IV) under grant 2022/04/Y/ST6/00001.